# Size dependent exciton dynamics in one-dimensional perylene bisimide aggregates

**Steffen Wolter[1], Janis Aizezers[1], Franziska Fennel[1], Marcus Seidel[1], Frank Würthner[2], Oliver Kühn[1] and Stefan Lochbrunner[1]**

[1] Institut für Physik, Universität Rostock, Universitätsplatz 3, 18055 Rostock, Germany

[2] Institut für Organische Chemie and Röntgen Research Center for Complex Material Systems, Universität Würzburg, Am Hubland, 97074 Würzburg, Germany

E-mail: stefan.lochbrunner@uni-rostock.de

**Abstract.** The size dependent exciton dynamics of one-dimensional aggregates of substituted perylene bisimides are studied by ultrafast transient absorption spectroscopy and kinetic Monte-Carlo simulations in dependence on the temperature and the excitation density. For low temperatures the aggregates can be treated as infinite chains and the dynamics is dominated by diffusion driven exciton-exciton annihilation. With increasing temperature the aggregates decompose into small fragments consisting of very few monomers. This scenario is also supported by the time dependent anisotropy deduced from polarization dependent experiments.

## 1. Introduction

In future, supramolecular organic structures will play an important role in photonic applications like organic solar cells [1] or optoelectronics [2, 3]. In these devices the behavior of excitons is crucial for their performance [4, 5]. Aggregates of chromophores [6] are attractive building blocks for such applications since they can host mobile excitons and energy can be transferred along them via exciton diffusion [7, 8]. Recently, we showed that excitons on aggregates of the perylene bisimide *N*,*N*'-di[*N*-(2-aminoethyl)-3,4,5-tris(dodecyloxy)benzamide]-1,6,7,12-tetra(4-*tert*-butylphenoxy)-perylene-3,4:9,10-tetracarboxylic acid bisimide (**PBI-1**, see Scheme 1) in methylcyclohexane (MCH) exhibit a fairly high mobility along one dimension [9]. This is in agreement with the notion that the monomers in these **PBI-1** aggregates are arranged in column like structures [10]. Hydrogen bonds between NH- and carboxyl groups of neighboring monomers and the interaction of the π-systems result in a π-stacked arrangement of the chromophores with a small lateral displacement. The angle between the monomeric transition dipoles and the columnar axis is somewhat smaller than the magic angle of 54.7°. Accordingly **PBI-1** aggregates are of J-type with a comparable small width of the exciton band [11]. The flexibility of the columns causes a significant degree of energetic disorder [12, 13] and restricts the coherence length of the excitons to only about two monomers [9].

To investigate the behavior and mobility of the excitons and their interaction with each other we study the time dependent exciton population and the dynamics induced by exciton-exciton annihilation with ultrafast pump-probe absorption spectroscopy. Our previous studies were restricted to very long aggregates, which were treated as infinite chains [9]. This poses the question under what circumstances this assumption is justified and what happens in case of short aggregates. Here we apply our methodology to aggregates with different lengths and compare our results to a hopping model which uses the kinetic Monte-Carlo technique and can treat chains of finite length.

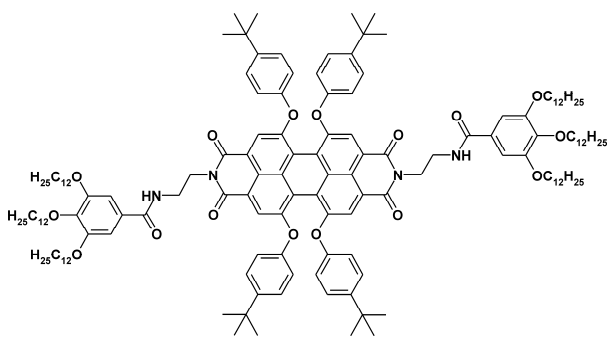

**Scheme 1.** Structure formula of the investigated dye (**PBI-1**). The aggregates are stabilized by the interaction between the π-systems of the central perylene chromophore of adjacent monomers and the formation of hydrogen bonds between oxygen atoms and the NH-groups of the terminal substituents.

Concentrations above $10^{-5}$ M of **PBI-1** in MCH are used, as the dye forms J-aggregates at room temperature in this concentration regime but dissociates at higher temperatures. This can be seen from the temperature dependent absorption spectra shown in Fig. 1 for a **PBI-1** solution in MCH with a concentration of 0.2 mM. At low temperatures the first absorption band is narrower, more intense and red shifted compared to the corresponding band at high temperatures like it is expected for the lowest excitonic transition of a J-aggregate compared to its counterpart in the monomer [14]. Accordingly, the temperature is a suitable parameter to control the size of the aggregates [15, 16].

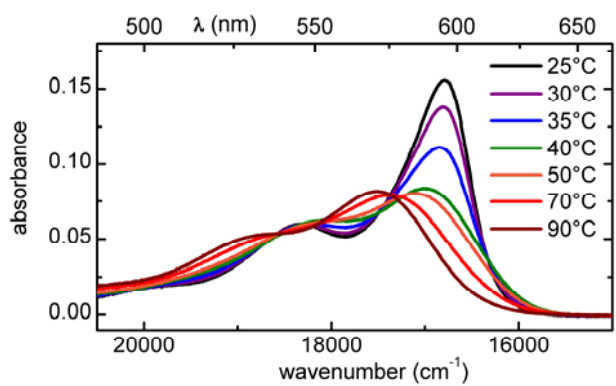

**Figure 1.** Absorption of **PBI-1** in MCH (c = 0.2 mM) for different temperatures.

In the following the applied experimental techniques and modeling approaches are introduced and then temperature dependent ultrafast transient absorption experiments are presented and compared to Monte-Carlo simulations. From the results we conclude on the mobility and interaction of the excitons and the decomposition behavior of the aggregates.

## 2. Experimental

The transient absorption measurements are performed with a pump-probe setup equipped with a non-collinearly phase matched parametric amplifier (NOPA) [17] delivering excitation pulses at 530 nm with pulse durations of 35 fs after compression with a prism sequence. A white light continuum [18] generated in a 4 mm thick calcium fluoride substrate [19] serves as probe and is dispersed after the sample by a fused silica prism and spectrally resolved detected by a photodiode array. The setup is pumped by a regenerative Ti:sapphire laser system (CPA 2001; Clark MXR) which provides ultrashort pulses at a wavelength of 775 nm and a repetition rate of 1 kHz. A combination of two wire-grid polarizers and an achromatic half wave plate allows to adjust independently the energy and polarization of the excitation pulses. The delay time between pump and probe is scanned with a motorized linear stage equipped with a retroreflector in the beam path of the pump light. The excitation pulses are focused to a spot diameter of 220 μm at the sample while the probe diameter is two times smaller there.

The sample solution is filled into a fused silica cell with a layer thickness of 1 mm. The cell is placed in a matching copper block which is electrically heated and which has two holes for the entering and exiting laser beams. Two thermo elements are used to measure the temperature and to provide a feedback signal for the temperature control unit. The temperature can be determined with an accuracy of 2.5°C.

## 3. Theoretical Models

*3.1 Diffusion Approach*

By integrating a suitable spectral interval of the transient spectra, as it will be described below, one obtains a signal, which is proportional to the density of the excitons. In a first approach to understand the decay dynamics of this signal at different temperatures, it is assumed, that it can be decomposed into two main contributions. The first contribution is caused by the mainly annihilation-driven decay of the excitons on the aggregates, which can be described by the differential equation Equ. 1 [7]:

$$\frac{dn(t)}{dt} = -\frac{n(t)}{\tau_1} - \frac{1}{2}\gamma(t)n^2(t) \tag{1}$$

$n(t)$ is the time-dependent exciton density, $\tau_1$ the intrinsic lifetime of the excitons and $\gamma(t)$ the annihilation rate. The factor of ½ within the second term takes into account, that only one exciton survives the annihilation process. According to our previous work on **PBI-1** aggregates, the exciton motion on the molecular chains is diffusive and restricted to one dimension. In this case $\gamma(t)$ depends on the diffusion constant $D$, the lattice constant $a$ and the molecule density $N_0$ according to Equ. 2:

$$\gamma(t) = \frac{1}{aN_0}\sqrt{\frac{8D}{\pi t}} \tag{2}$$

Equations 1 and 2 hold strictly only for infinite chains. Integration of Equ. 1 leads to [20]:

$$n(t) = \frac{n_0 \cdot \exp(-t/\tau_1)}{1 + n_0/aN_0\sqrt{2D\tau} \cdot \text{erf}(\sqrt{t/\tau_1})} \qquad (3)$$

$n_0$ is the initial exciton density in the sample directly after the pump pulse. The second contribution to the signal is caused by the radiative decay of excited **PBI-1** monomers or very small aggregates, which appear at higher temperatures and do not experience exciton-exciton annihilation since they host only one excitation. Their de-excitation is described by a single exponential decay with a time constant $\tau_2$. Thus, the full time-evolution of the integrated optical density $I(t)$ reads

$$I(t) = a_1 \cdot \frac{\exp(-t/\tau_1)}{1 + <n^*>/a\sqrt{2D\tau_1} \cdot \text{erf}(\sqrt{t/\tau_1})} + a_2 \cdot \exp(-t/\tau_2) \qquad (4)$$

The scaling factors $a_1$ and $a_2$ weight the two contributions at a given temperature and $<n^*>$ denotes the number of excited excitons per molecule.

*3.2 Kinetic Monte Carlo Approach*
In order to take the finite size of the one-dimensional aggregates into account kinetic Monte Carlo simulations [21] are used to model the exciton kinetics. An ensemble of aggregates is constructed with a total number of considered molecules $N_{mol}$ by equally partitioning them into $N_{agg}$ linear aggregates. Initially, $N_{ex}$ molecules are excited randomly according to the fraction $N_{ex}/N_{mol} = <n^*>$ of optically excited molecules given by the experiment. In the subsequent kinetics the effects of exciton hopping, exciton annihilation and radiative decay are taken into account by respective rates. For the exciton hopping between neighboring sites we use the hopping rate assuming Förster transfer. The rate $k_{mn}$ from site $m$ to site $n$ is given in the slow modulation limit by [22],

$$k_{mn} = \frac{2\pi}{\hbar} \frac{|J_{mn}|^2}{\sqrt{2\pi k_B T S_{mn}}} \exp\left[-\frac{(E_m - E_n - S_{mn}/2)^2}{2k_B T S_{mn}}\right] \qquad (5)$$

Here, $E_m$ denotes the site energies, which in principle could be different for different sites due to energetic disorder. In the here considered case energetic disorder is neglected, i.e. the site energies can be set to zero. $S_{mn} = S_m + S_n$ is the combined Stokes shift for the transitions at site $m$ and $n$ and $J_{mn}$ is the Coulomb coupling between the transitions. The monomer Stokes shift is $S_m = 1150$ cm$^{-1}$ [11] and the Coulomb coupling $J_{mn}$ will be treated as a parameter when fitting the experimental kinetics; only nearest neighbor coupling is assumed. Exciton-exciton annihilation is considered by introducing the bimolecular rate $k_{ann}$ which applies whenever two excitons are on the same site. Here a fixed value of $k_{ann} = 10$ ps$^{-1}$ is used. Finally, the population of each site is depleted according to a radiative decay rate $k_{rad}$ of 0.27 ns$^{-1}$ [9].

In the kinetic Monte Carlo simulations we have used 10000 realizations for the initial conditions according to $N_{ex}$. Specifically, we have used $N_{mol} = 2000$ and three concentration, i.e. $N_{ex} = 58, 110, 178$ corresponding to excitation densities $<n^*>$ of 0.029, 0.055, and 0.089, respectively. Note, that in case the number of molecules was not a multiple of the aggregate size the nearest integer value for $N_{mol}$ has been used. Further, we excluded initial double excitation of a site. During the simulation the total exciton population with respect to all molecules is monitored. Coulomb coupling and aggregate size are systematically varied such as to obtain the best fit to the experimental data.

After obtaining an optimal set of parameters for fitting the experimental decay curves, we have extended the model to include interchain hopping with a rate $k_{int}$. This allows us to scrutinize the initial assumption of strictly one-dimensional exciton transport.

**4. Experimental Results**

Figure 2 shows transient spectra measured after optical excitation of **PBI-1** dissolved in MCH with a concentration of 0.2 mM for a temperature of 30°C and of 90°C at five different delay times. The energy per pump pulse is 120 nJ and pump and probe polarization are set to the magic angle 54.7° between each other. In this way pump induced anisotropies do not influence the transient absorption signal. Further measurements were carried out at intermediate temperatures. In all cases a strong negative optical density change is observed which can be attributed to a combination of ground state bleach and stimulated emission from the populated electronic state. The spectral widths and positions correlate nicely with the ground state absorption at the respective temperature (see Fig. 1). The excited state absorption (ESA) is overlaid by the ground state bleach and the stimulated emission, e.g. for the long aggregates the ESA leads to an almost zero differential optical density in the range below 575 nm.

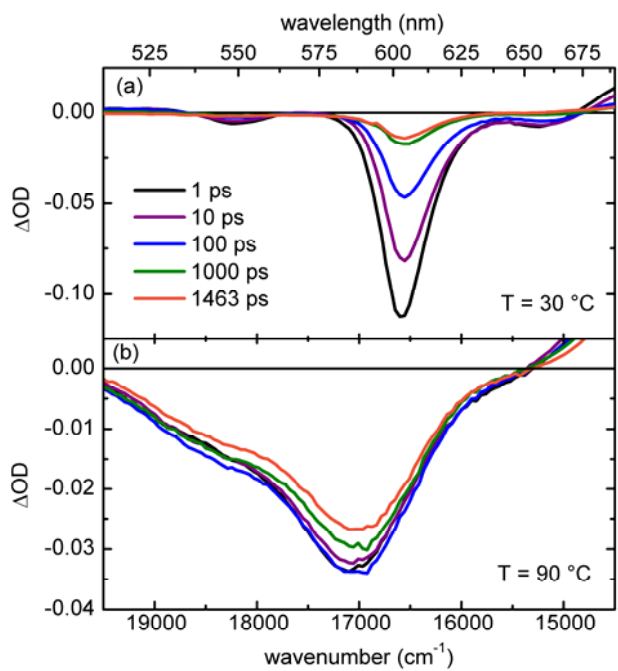

**Figure 2.** Transient absorption spectra for different delay times after optical excitation of 0.2 mM **PBI-1** dissolved in MCH, (a) at 30°C and (b) at 90°C.

At a temperature of 30°C the negative signal decays non-exponentially on a time scale of a few hundred picoseconds while at 90°C the signal decay is much less pronounced and occurs mostly after one nanosecond. The changes in the spectral shape of the transient spectra are in both cases small indicating that the signal evolution is dominated by a simple relaxation of the population of the electronically excited $S_1$ state back to the ground state. To obtain a measure for this population, i.e. the number of excitons, and to eliminate the influence of the small spectral changes the negative band of the differential optical density is integrated. The time dependent integrated optical density is shown in Fig. 3a for three different temperatures.

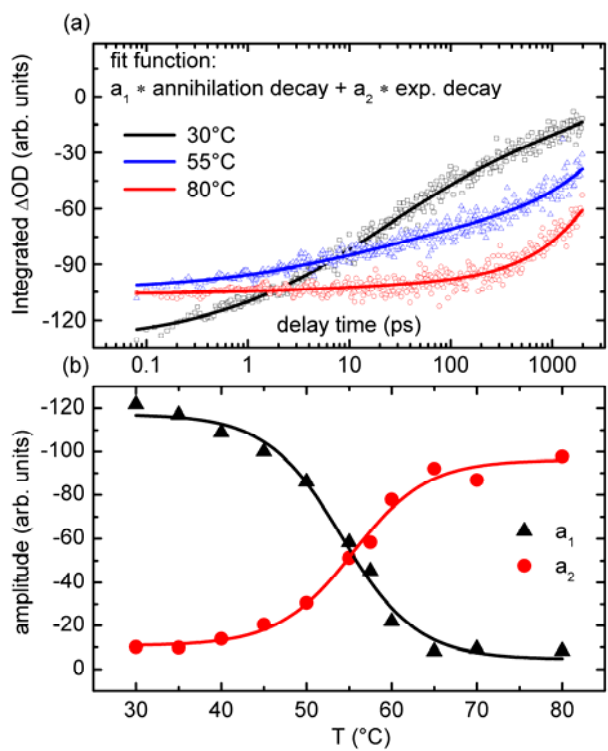

**Figure 3.** (a) Integrated transient absorption signal due to optical excitation of **PBI-1** in MCH (c = 0.2 mM, $<n^*>$ = 0.055, see text) for three different temperatures (symbols). The solid lines are model functions fitted to the data, which consist of two contributions, exciton-exciton annihilation and a single exponential decay. (b) Amplitudes of the two contributions in dependence of the temperature. Black refers to exciton-exciton annihilation and red to the single exponential decay. The solid lines are guides to the eye.

To analyze the behavior of the integrated signal, Equ. 3 is fitted to the data (solid curves in Fig. 3a). The fraction of initially excited molecules in the sample $<n^*> = n_0 / N_0$ can be obtained from the experimental data as described in [9]. Taking saturation effects into account results in an excitation density $<n^*>$ of 0.055 per monomer for the applied pulse energy. The lifetime of the excitons on the aggregates $\tau_1 = 3.6$ ns and of the monomers $\tau_2 = 4.1$ ns are known from our previous work [9] and the lattice constant is assumed to be $a = 0.48$ nm [10]. Accordingly, the scaling factors $a_1$ and $a_2$ and the diffusion constant $D$ are the only free parameters. If we assume, that at low temperatures, almost all PBI molecules are integrated in aggregates, the diffusion constant for the long J aggregates can be determined to be $D = (0.9\pm0.3)$ nm²/ps, leaving the scaling factors as only free parameters. Therefore, the fractions of monomers and aggregates in the sample can be obtained. At this point it is important to note, that the J-aggregates are still treated as infinite chains within this model. Accordingly, we cannot make any prediction about the actual aggregate size at a given temperature. The found diffusion constant deviates somewhat from the one reported in [9]. This is probably due to uncertainties in determining the initial exciton concentrations and the influence of the exponential contribution, which does not appear in the original model.

Figure 3a shows that the model described by Equ. 3 can reproduce the exciton dynamics for all temperatures very well. At low temperatures, the annihilation driven decay dominates the dynamics, indicating, that most of the dye molecules are aggregated (amplitude $a_1$ in Fig. 3b). As the temperature rises, amplitude $a_1$ declines whereas the contribution of the radiative decay of the monomers $a_2$ rises. The change from annihilation dominated dynamics to a single exponential decay occurs within a temperature window of about 25 K around 55°C. Therefore one may conclude that upon heating the **PBI-1** aggregates brake up into monomers without undergoing any intermediate state.

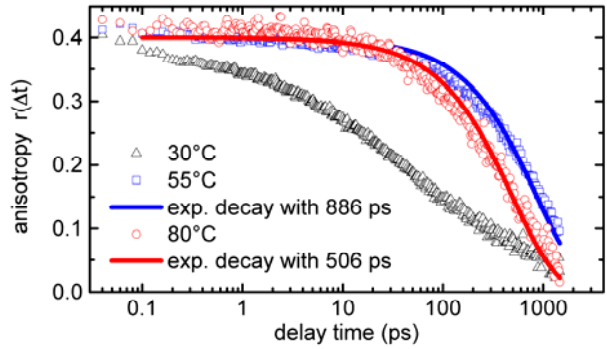

**Figure 4.** Time evolution of the anisotropy-parameter $r(t)$ for different temperatures. The curves for 55°C and 80°C are fitted with a single exponential decay.

To test this notion, polarization-dependent measurements were carried out. A common measure for the polarization dependence is the transient anisotropy parameter $r(t)$ that can be obtained by changing the relative orientation of the pump and probe polarization from parallel polarizations yielding the signal $Sig_{\parallel}(t)$ to perpendicular polarizations yielding $Sig_{\perp}(t)$ [23, 24]:

$$r(t) = \frac{Sig_{\parallel}(t) - Sig_{\perp}(t)}{Sig_{\parallel}(t) + 2Sig_{\perp}(t)} \quad (5)$$

At low temperatures, a fast decay of the transient anisotropy during the first 20 picoseconds is observed (see Figure 4). This fast decay is caused by the exciton-exciton-annihilation on the J-aggregates, that are present at this temperature. Due to the strong excitation energy dependence of the annihilation probability [9], the signal for parallel polarized pump and probe beam $Sig_{\parallel}(t)$ decays much faster than $Sig_{\perp}(t)$, leading to an accelerated decay of the transient anisotropy. Interestingly, this fast decay dynamics slows strongly down at intermediate temperatures where also the single exponential decay dynamics replaces annihilation. This implies that the excitons loose their mobility and the influence of annihilation on the anisotropy reduces drastically. Furthermore, we see that the transient anisotropy decays slightly faster at high temperatures than at intermediate temperatures. If we assume, that the anisotropy decay in both cases is mainly caused by rotational diffusion of the excited species, the dynamics can be modeled by a single exponential decay starting at a value of $r(t=0) = 0.4$ [25]:

$$r(t) = 0.4 \cdot \exp(-t/\tau_{rot}) \tag{6}$$

$\tau_{rot}$ is the time constant for the rotational diffusion process that is connected to the hydrodynamic volume $V$ of the rotating species by the Stokes-Einstein-relation [25]:

$$\tau_{rot} = \frac{V\eta(T)}{k_B T} \tag{7}$$

$\eta(T)$ is the temperature dependent viscosity of the solvent MCH and $k_B$ is the Boltzmann constant. Figure 4 shows that the rotational diffusion model is in good agreement with the measured data for high as well as for intermediate temperatures. Calculating the hydrodynamic volume of the rotating species allows us to make at least qualitative statements upon the size of the species. For a temperature of 55°C a value of $V = (9.43\pm0.22)$ nm³ and for 80°C a value of $V = (7.12\pm0.16)$ nm³ is calculated, indicating, that the rotating species at intermediate temperatures is not a monomer but a very small PBI aggregate, probably a dimer. In this case it is unlikely that two excitons reside on the same aggregate and annihilation plays no role.

A crude approximation for the hydrodynamic volume of the rotating species can be done by assessing the size of the PBI molecules from the density $\rho = 1.5$ g/cm³ of comparable perylene bisimides given in [26]. The single molecule volume can be estimated by

$$V = \frac{M}{\rho \cdot N_A} \tag{8}$$

where $M = 2383$ g/mol [10] is the molar mass of PBI and $N_A$ the Avogadro constant. Equ. 8 yields a single molecule volume of $V = 2.64$ nm³, but it has to be mentioned, that the measured hydrodynamic volumes include an additional solvent shell. For simplicity, the **PBI-1** molecules are treated as spheres. To account for the solvent shell, the radius of the single **PBI-1** hard sphere is increased by the thickness $r_{MCH} = 0.38$ nm of one MCH monolayer, that is calculated from the structure of the MCH molecule. Recalculating the volume of the whole sphere including the shell we arrive at a volume of 7.94 nm³ which is in proximity to the hydrodynamic volume of 7.12 nm³ measured at 80°C. To calculate the hydrodynamic volume of a small aggregate, we repeat the above procedure after replacing $V$ by $N \cdot V$ for an aggregate consisting of $N$ molecules. Thereby, we arrive at a volume of 13.04 nm³ for the **PBI-1** dimer which is close to the hydrodynamic volume of 9.43 nm³ measured at 55°C. This consideration seems to approve our above statement, that at high temperatures the sample mainly consists of **PBI-1** monomers whereas at intermediate temperatures very small aggregates are present.

## 5. Comparison with Hopping Model

While Equ. 3 is strictly valid only for an infinite chain the average aggregate length is expected to shorten with increasing temperature [15]. The limited size should influence the dynamics since the number of excitons, which can interact with each other, reduces with decreasing aggregate length. To quantify this effect and to test the applicability of the diffusion model a hopping model based on a Monte Carlo approach is used to simulate the exciton dynamics. In Fig. 5 the measured exciton populations are compared to Monte Carlo simulations and the analytic diffusion model for three different excitation densities between 2.9 % and 8.9 % of the chromophores. The fitting has been performed for an excitation density of 2.9 % which gave using the parameters from section 3.2 a value for the coupling matrix element of $J_{mn} = 120$ cm$^{-1}$, which is in good agreement with our previously measured value of 110 cm$^{-1}$ [9]. Notice, however, that this value has to be considered as being an effective coupling since it reflects the underlying hopping model, i.e. either Equ. (5) or the diffusing equation. The bare electronic Coulomb coupling has been calculated to be -513 cm$^{-1}$ in Ref. [27]. Furthermore, we found that beyond an aggregate size of 80 monomers there is little size dependence of the kinetics. The thus obtained parameter set is used for the higher excitation densities as well.

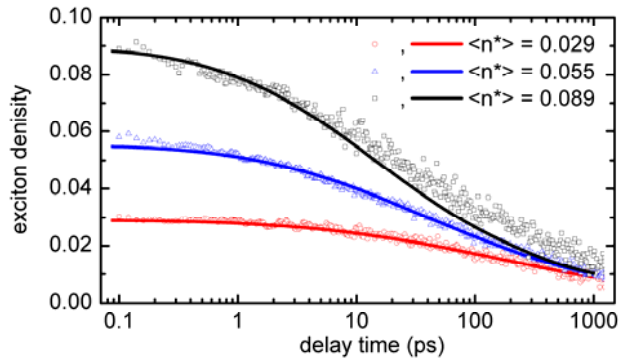

**Figure 5.** Comparison between the hopping model (lines) and the experimental data at 30°C (symbols) for different initial excitation densities $<n^*>$.

The hopping rate $k_{hopp} = 2 \cdot k_{mn} = 5$ ps$^{-1}$ calculated by Equ. 5 can be compared to the experiment. The diffusion constant

$$D = \frac{1}{2} a^2 k_{hopp} \qquad (9)$$

is directly related to the hopping rate and the distance $a$ between two hopping sites. If the distance is taken as the monomer separation (see above) the fitted annihilation model corresponds to a hopping rate $k_{hopp} = (7.8 \pm 2.7)$ ps$^{-1}$. This is in reasonable agreement with the Monte-Carlo simulations.

The deviations between the Monte-Carlo simulations and the analytic model increase as expected with decreasing aggregate length (see Fig. 6a). If the number of hopping sites is less than 45 the differences are so pronounced that they should be clearly seen in the experiment. From this we conclude that the number of intermediate aggregates, which show significant annihilation but cannot be treated as infinite chains, is low. In other words the long aggregates encountered at low temperatures dissociate with increasing temperature into small fragments containing at most a few monomers.

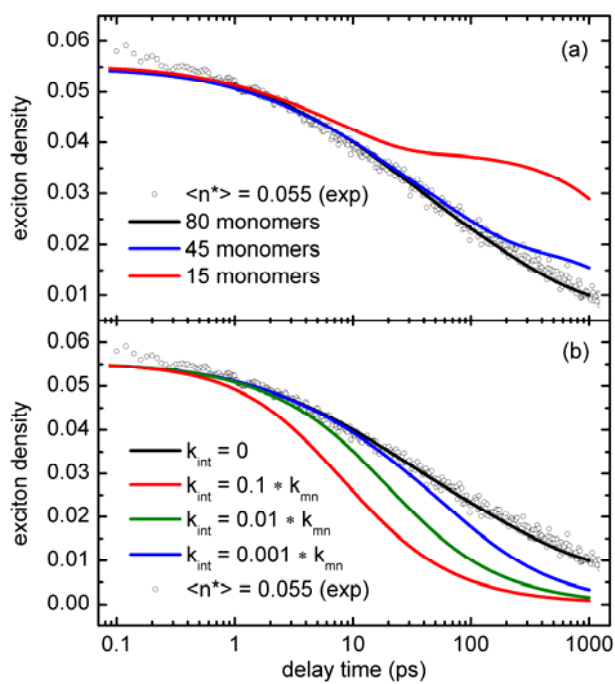

**Figure 6.** Experimental data for $<n^*> = 0.055$ at 30°C compared to (a) the hopping model for different aggregate sizes and to (b) the hopping model with an additional interchain hopping rate $k_{int}$.

The results of the extended two-dimensional Monte-Carlo model with an interchain hopping rate $k_{int}$, given a fraction of the optimal intrachain rate $k_{mn}$ are shown in Fig 6b for the case of intermediate excitation density. Apparently, including interchain hopping leads to strong deviations from the observed kinetics. Even if the jump rate between aggregate chains is several orders of magnitude smaller than the hopping rate along the aggregates a significant discrepancy to the experimental results emerges. This observation gives evidence that the excitons cannot jump between the aggregates and the aggregates are well described as one-dimensional chains along which the excitons can move.

**Summary and Conclusions**
The size dependent exciton dynamics of one-dimensional **PBI-1** aggregates is studied by ultrafast transient absorptions spectroscopy and kinetic Monte-Carlo simulations in dependence on the temperature and the excitation density. For low temperatures the aggregates can be treated as infinite one-dimensional chains and the dynamics is dominated by diffusion driven exciton-exciton annihilation. With increasing temperature the aggregates decompose into small fragments consisting of very few monomers. The chances that two excitons on one of such fragments are excited are very low and annihilation processes cannot occur anymore. Aggregates of intermediate size seem not to contribute significantly. Polarization dependent experiments support this scenario since the anisotropy decay at intermediate temperatures points to orientational relaxation of entities with a volume of several monomers. At high temperatures the anisotropy decay accelerates somewhat indicating an increasing fraction of monomers.

**Acknowledgement**

Financial support by the German Science Foundation via the collaborative research center SFB 652 "Strong correlations and collective effects in radiation fields: Coulomb systems, clusters and particles" is gratefully acknowledged.

**References**

[1] Clarke T M and Durrant J R, *Chem. Rev.*, 2010, **110**, 6736–6767.

[2] Zhan X, Facchetti A, Barlow S, Marks T J, Ratner M A, Wasielewski M R and Marder S R, *Adv. Mater.*, 2011, **23**, 268–284.

[3] Martens S C, Zschieschang U, Wadepohl H, Klauk H and Gade L H, *Chem. Eur. J.*, 2012, **18**, 3498–3509.

[4] Qu J, Zhang J, Grimsdale A C, Müllen K, Jaiser F, Yang X and Neher D, *Macromolecules*, 2004, **37**, 8297–8306.

[5] Armstrong N R, Wang W, Alloway D M, Placencia D, Ratcliff E and Brumbach M, *Macromol. Rapid Commun.*, 2009, **30**, 717–731.

[6] Würthner F, Kaiser T E and Saha-Möller C R, *Angew. Chem. Int. Ed.*, 2011, **50**, 3376–3410.

[7] Engel E, Leo K and Hoffmann M, *Chem. Phys.*, 2006, **325**, 170 – 177.

[8] Möbius D, *Adv. Mater.*, 1995, **7**, 437–444.

[9] Marciniak H, Li X-Q, Würthner F and Lochbrunner S, *J. Phys. Chem. A*, 2011, **115**, 648–654.

[10] Li X-Q, Zhang X, Ghosh S and Würthner F, *Chem. Eur. J.*, 2008, **14**, 8074–8078.

[11] Ambrosek D, Marciniak H, Lochbrunner S, Tatchen J, Li X-Q, Würthner F and Kühn O, *Phys. Chem. Chem. Phys.*, 2011, **13**, 17649–17657.

[12] Fidder H, Knoester J and Wiersma D A, *J. Chem. Phys.*, 1991, **95**, 7880–7890.

[13] Knapp E W, *Chem. Phys.*, 1984, **85**, 73–82.

[14] Spano F C, *Acc. Chem. Res.*, 2010, **43**, 429–439.

[15] Chen Z, Stepanenko V, Dehm V, Prins P, Siebbeles L D A, Seibt J, Marquetand P, Engel V and Würthner F, *Chem. Eur. J.*, 2007, **13**, 436–449.

[16] Tian Y, Stepanenko V, Kaiser T E, Würthner F and Scheblykin I G, *Nanoscale*, 2012, **4**, 218–223.

[17]   Wilhelm T, Piel J and Riedle E, *Opt. Lett.*, 1997, **22**, 1494–1496.

[18]   Alfano R R and Shapiro S L, *Phys. Rev. Lett.*, 1970, **24**, 584–587.

[19]   Nagura C, Suda A, Kawano H, Obara M and Midorikawa K, *Appl. Opt.*, 2002, **41**, 3735–3742.

[20]   Gulbinas V, Chachisvilis M, Valkunas L and Sundström V, *J. Phys. Chem.*, 1996, **100**, 2213–2219.

[21]   Kalos M H and Whitlock P A, *Monte Carlo Methods*, Wiley-VCH, Weinheim, 2008.

[22]   May V and Kühn O, *Charge and Energy Transfer Dynamics in Molecular Systems*, 3$^{rd}$ edition, Wiley-VCH, Weinheim, 2011.

[23]   Fleming G, Morris J and Robinson G, *Chem. Phys.*, 1976, **17**, 91–100.

[24]   Lessing H and Jena A, *Chem. Phys Lett.*, 1976, **42**, 213–217.

[25]   Valeur B, *Molecular Fluorescence: Principles and Applications*, Wiley-VCH, Weinheim, 2002.

[26]   Hädicke E and Graser F, *Acta Cryst.,* 1986, **C42**, 189-195 .

[27]   Ambrosek D and Kühn O, *in preparation*.